# Considering functional spreadsheet operator usage suggests the value of Example Driven Modelling for Decision Support Systems

S. Thorne, D. Ball
University of Wales Institute Cardiff
Sthorne@uwic.ac.uk Dball@uwic.ac.uk

**Abstract**

Most spreadsheet surveys both for reporting use and error focus on the practical application of the spreadsheet in a particular industry. Typically these studies will illustrate that a particular percentage of spreadsheets are used for optimisation and a further percentage are used for 'What if' analysis. Much less common is examining the classes of function, as defined by the vendor, used by modellers to build their spreadsheet models. This alternative analysis allows further insight into the programming nature of spreadsheets and may assist researchers in targeting particular structures in spreadsheet software for further investigation. Further, understanding the functional make-up of spreadsheets allows effective evaluation of novel approaches from a programming point of view. It allows greater insight into studies that report *what* spreadsheets are used for since it is explicit which functional structures are in use in spreadsheets. We conclude that a deeper understanding of the use of operators and the operator's relationship to error would provide fresh insight into the spreadsheet error problem. Considering functional spreadsheet operator usage suggests the value of Example Driven Modelling for Decision Support Systems

**Electronic Spreadsheets and business**

Research has shown that the use of spreadsheets in organisations is pervasive, used in many industries and for many purposes. This pervasive use of spreadsheets is reflected in a gradual increase in spreadsheet error research; this is partly due to the slow realisation that spreadsheets, as end user tools, are error prone and that spreadsheets are used in strategic business applications

Most End User Development (EUD) academic surveys such as (Davies 1987, Jenne 1996 and Taylor *et al.* 1998) rarely focus on spreadsheets. It is therefore difficult to precisely say *who* uses spreadsheets and for what purpose. However, there are some surveys that provide data on spreadsheet usage specifically. Pemberton and Robson (2000) surveyed 227 respondents from a mix of Private and Public sector organisations in the UK. They found that over 80% of respondents were spreadsheet users at some level. The activities spreadsheets were used for ranged from data storing to Decision Support Systems (DSS) implemented in spreadsheets.

The Spreadsheet Engineering Research Programme (SERP) is an international research effort aimed at surveying spreadsheet usage across a number of large organizations. It is not currently complete but has already surveyed over 1300 participants in a number of organizations in the United States and the United Kingdom. SERP (2005) found that spreadsheets are used predominantly for alternative modeling (what if analysis) with 98% of respondents indicating so. This was closely followed by 90% using spreadsheets for data analysis. The most common technique used in these models is statistical analysis (49%) followed by optimization techniques (31%).

Interestingly the results presented by SERP (2005) differ significantly to Pemberton and Robson (2000). The SERP data suggests that most spreadsheets are used for "what if" analysis using statistical techniques, conversely Pemberton and Robson (2000) found that statistical analysis was the least popular use for spreadsheets, the most popular being data sorting.

Further, Brancheau and Wetherbe (1990) found that the adoption of spreadsheet applications by users started in 1981 with VisiCalc and grew significantly until 1987 when the survey was completed. In historical terms 1987 marked the introduction of GUIs and subsequently excel was introduced in the same year. Taylor *et al.* (1998) surveyed 34 UK organisations and found that 85% of those organisations used spreadsheets as their primary method for EUD. However, since this study was aimed at EUD and not spreadsheets, there is no further detail on type of use.

In addition, there are case studies and audit experiences written by both academics and practitioners. Such investigations are usually driven by some financial loss or realisation that the practice they are partaking in carries some significant risk. Two such cases are Fernandez (2002) and Gosling (2003) who both surveyed spreadsheet usage and policy in their respective organisations. Both found that spreadsheets are used as low-level data stores and for calculation of trivial items such as expenses. They also discovered that spreadsheets had become part of the IT infrastructure of the organisations and that the business would be seriously affected if these spreadsheet applications failed. This clearly shows a strategic reliance on spreadsheets for decision analysis.

Croll (2005) reports on the usage of spreadsheets in the financial markets in London. Croll found that there is great reliance on spreadsheets for modelling as the below quote from one of the participant responses highlights.

*"Excel is utterly pervasive. Nothing large (good or bad) happens without it passing at some time through Excel"* (Croll, 2005)

The evidence provided so far in this section suggests spreadsheets as the most commonly used end user tool. There is also evidence to suggest that spreadsheets are used in a strategic manner (Croll 2005, Fernandez 2002, Gosling 2003). Croll's evidence comes in the form of quotes from participants studied. Both Fernandez and Gosling are vertical case studies in two different organisations. Fernandez investigates an international private company and Gosling a large National Health Service (NHS) trust. The aim of both studies was to establish the use of spreadsheets in each organisation and examine the implications of their use. Both Fernandez and Gosling found that spreadsheets were pervasive in the organisations and that spreadsheets were used in a strategic manner. In some cases spreadsheets were used to make decisions on how an entire department was run. Spreadsheets had also become part of the Information Systems Architecture, removing data from the corporate system, manipulating it and then re entering it into the Corporate system. Clearly this practice of merging validated and un-validated data is undermines the integrity of corporate systems.

**Spreadsheet Errors**

The first study into spreadsheet error was conducted by Brown and Gould for IBM in 1987. This study took 9 experienced spreadsheet developers and examined their performance when asked to create a number of spreadsheets from scratch. They found that 44% of the spreadsheets developed contained errors such as mistyping formulae. This study was conducted because the authors had noted that business spreadsheet usage had boomed and that it had been suggested that spreadsheets might contain errors.

Since this original paper, there have been many studies in spreadsheet error and the statistics reported from these studies varies from 30% to 100% of models with errors. Table 1 depicts some experimental studies with relevant error rates.

| **Author and Year** | **Percentage of models with errors** |
|---|---|
| Hicks and Panko, 1995 | 91% |
| Javrin and Morrison, 1996 | 84% |
| KPMG, 1997 | 91% |
| Panko and Halverson, 1997 | 80% |
| Javrin and Morrison, 2000 | 95% |

**Table 1 Spreadsheet error rates**

These statistics have led other researchers to investigate spreadsheet error in more depth, asking a variety of pertinent questions: what causes errors, how many types are there, how can errors be reduced or removed. Thus far most spreadsheet error research studies have considered error from a general point of view, i.e. a focus on the number of models produced with error or the average number of errors per cell. Whilst this information is useful for estimating the extent of error in certain domains it does not necessarily tell us why error comes about or what causes the error. Research into taxonomies of error such as (Panko and Halverson 1998, Teo and Tan 1999 and Ayalew et al 2000, Rajalingham *et al.*

2000, Rajalingham 2005) have all examined and defined error types in varying levels of detail. These error types are not specific to the actual application, rather they are generic.

**Limitations of current spreadsheet research**

General surveys of spreadsheet error have traditionally focused on the end user products, i.e. the studies are themed according to the final application of the technology. They are often written from a management point of view, highlighting inadequacies in policy or practice and attributing these inadequacies to poor spreadsheet quality. Typical output would be the percentage of spreadsheets used in the accounting industry together with the percent used for optimisation. Whilst this serves a purpose, it does not shed any new light on the nature of spreadsheet error or what causes spreadsheets to be so error prone. An alternative view of reporting spreadsheet error is to examine the programming structures that spreadsheets are composed of. Programming structures in spreadsheets consist of formulae constructed utilising built in vendor operators. A deeper understanding of the use of operators and the operators relationship to error would provide fresh insight into the spreadsheet error problem. Within spreadsheet software, there are a number of vendor defined classes of function, each function contains various operators that relate to the class they are a member of.

**Functional Classes in Spreadsheet Software**

As defined by the vendor Microsoft, there are 11 classes of function offered with the standard Excel spreadsheet software. Excel is chosen since it is the most commonly used spreadsheet application according to Walchenbach (2005). Walchenbach states that Excel now accounts for 90% of the spreadsheet market, although it is difficult to determine the exact number of Excel users, in 1997 alone Microsoft shipped over 70 million copies of Excel 97.

These classes contain operators to be used in formulae expressions and are grouped according to their actual purpose. The 11 classes contain varying amounts of operators ranging from 5 to 78 operators in a class, offering a grand total 343 unique operators. The 11 class groupings are shown in table 2.

| Class Name | Number of operators |
|---|---|
| Database | 12 |
| Date and Time | 20 |
| Financial | 53 |
| Engineering | 39 |
| Information | 18 |
| Logical | 6 |
| Look-up and Reference | 17 |
| Math and Trigonometry | 60 |
| Statistical | 78 |
| Text | 35 |
| External linking | 5 |

**Table 2 Excel function classes**

**Studies of functional usage**

As previously mentioned, there have been very few documented studies of operator functionality usage. The studies that do exist offer some insight but often lack detail or are a minor aspect of a larger study. Chan and Storey (1996) surveyed 256 analysts using Lotus 123 on the functionality of spreadsheets used, see figure 1.

The survey was based upon a Likert scale (1 being never and 5 always).The participants indicated how often they use a particular class of function in their spreadsheet and that was recorded on a Likert scale. For example if they never used the Goal Seek function, this would be recorded as a 1 on the Likert scale. The main findings of this study, see figure 1, show that mathematical and statistical functions are the most frequently used and that goal seeking is the least used. However, since this study was conducted on Lotus 123 users, the functional classes are different to that of Excel. Unfortunately the vendor Lotus were unable to provide a detailed functionality listing for Lotus 1-2-3. The difference between Lotus and Excel makes direct comparison difficult, e.g. some operators in Excel are not supported in Lotus 1-2-3 and vice-versa.

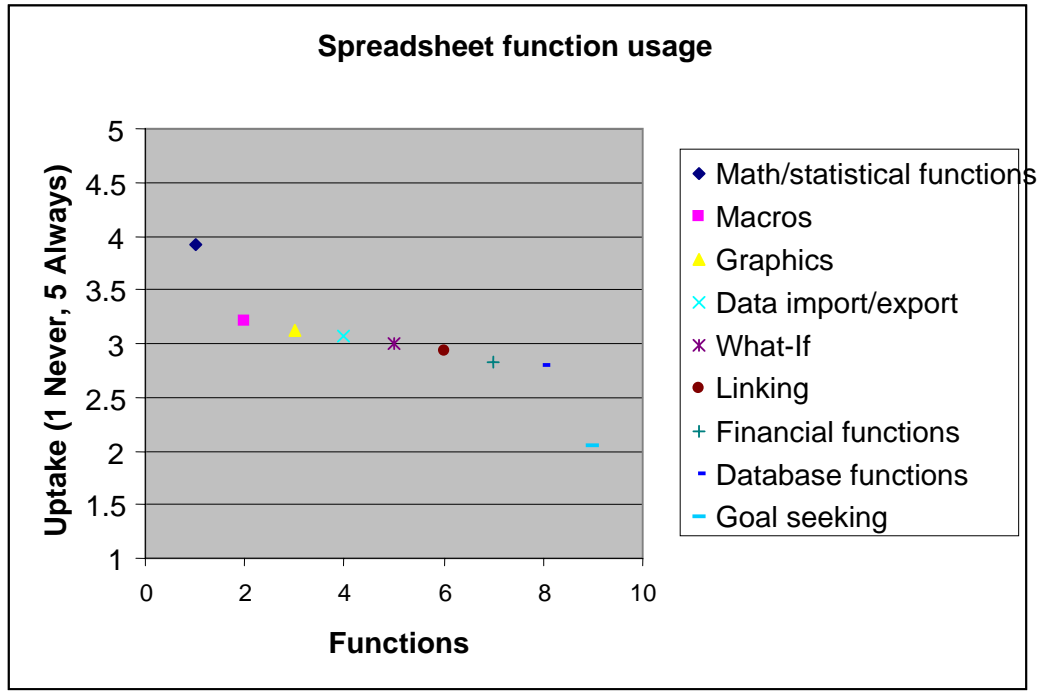

**Figure 1 Chan and Storey (1996) Frequency of spreadsheet operator use**

Ballinger *et al.* (2003) presented spreadsheet functional data collected from 259 Excel workbooks used to record student marks in a University. Figure 2 shows the results of the survey, in this case the data shows how many operators of a function type were used, i.e. there were 751 logical operators in the 259 workbooks.

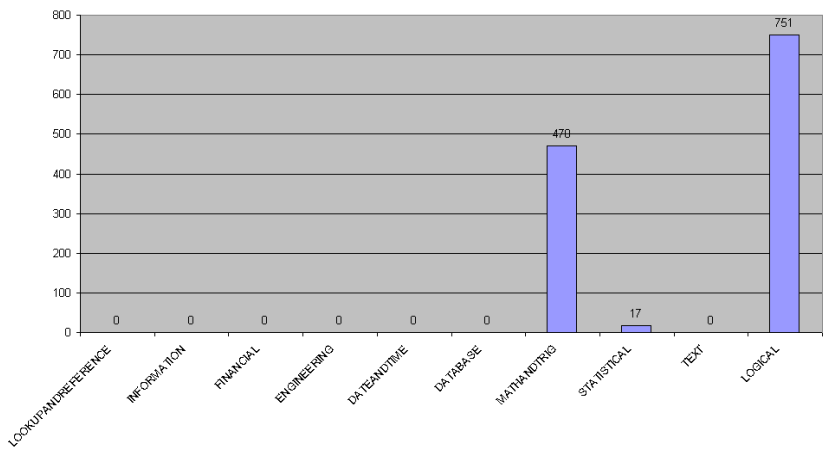

**Figure 2 (Ballinger *et al.*, 2003) Frequency of spreadsheet operator use**

The data clearly shows that logical and math functions are used more extensively than other classes. Indeed, most classes were not used in the worksheets which suggest that the sample may be biased due to the specific application.

The results of this study concur with that of Chan and Storey (1996) to some extent. Both studies identify that mathematical functions are used extensively. However, it is unclear if Chan and Storey (1996) include the Logic operators in their mathematics class.

Through Private Communication with Barry Lawson of The Tuck management school Dartmouth College in the US, further data regarding functional utilisation was obtained. The data was extracted from a 'base of knowledge' gathered at the college via the Spreadsheet Engineering Research Project (SERP). The data was based upon 35 randomly selected spreadsheets that were submitted by the schools alumni. The results of the study are presented in figure 3. Figure 3 suggests there is a disproportionate amount of Financial and Statistical function usage, given the other studies. One possible reason for this apparent bias is that the data was extracted from the Tuck schools alumni who had all studied financial management in some sense on predominantly MBA programmes which traditionally contain financial management aspects.

*"I would only observe that because we are a business school, one might expect that our contacts may be biased in that direction - hence perhaps a larger than random number of financial spreadsheets"*

Barry Lawson, SERP, Tuck School of Management (2006)

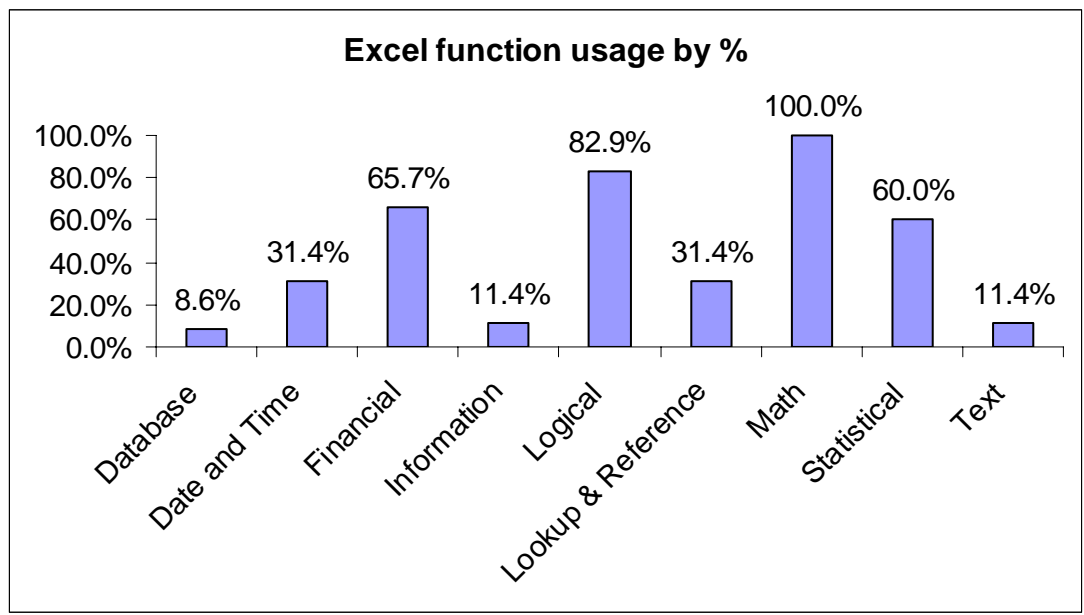

**Figure 3 (SERP, 2006) Spreadsheet function use by percentage**

The results of the data were arrived at by examining each of the 35 spreadsheets and determining what functions each spreadsheet contained. For example the Math class was used by 100% of spreadsheets in the survey, i.e. all 35 of the spreadsheets examined used one of more math functions. This does suggest some confusion in interpreting figure 3. For example if 65.7% of spreadsheets contain one example of a financial operator then we get the bar on the figure as shown. On the other hand if there are 100 examples of all 53 financial operators in 65.7% of spreadsheets then we would apparently get the same bar resulting. However, this study is more comprehensive than either of the previous studies, nevertheless it does share some commonality in the results. All three studies (SERP 2006, Ballinger 2003 and Chan and Storey 1996) identify that Math functions are used extensively in spreadsheets. Both SERP (2006) and Ballinger (2003) identify that Logical functions are used extensively. If we consider the possibility that Chan and Storey interrupt logical operators as part of the math class, this further reinforces this theory.

For further information regarding SERPs research visit
http://mba.tuck.dartmouth.edu/spreadsheet/index.html

**Conclusions of functional usage analysis**

From the data available one might hypothesise a research question: "Are the majority of functions used in spreadsheets either of the math or logical class?" Since the SERP (2006) data offers the most comprehensive data, it is the best indicator as to the proportions of function classes in spreadsheets. This data suggests that Math functions appear in near to 100% of spreadsheets and that Logic functions appear in around 80% of all spreadsheets, however this does not indicate how many of each type occur. One possibility is that modellers may prefer to build their own bespoke models via simpler operators rather than utilise some pre defined operator. Further, possibly a series of statements built with simpler operators may suggest an increased use of logical operator connectives.

This theory is supported by Napier (1989 and 1992) who demonstrated that not only did spreadsheet modellers use very little of the functionality in spreadsheets but were largely unaware of much of the functionality on offer. If we apply this theory to the rest of the data contained in this survey, it further supports the notion that simple spreadsheet operators are more useful than specific pre defined functions. This may explain why Ballinger *et al.* had a high number of Maths and Logic functions and no others except a minority using Statistical functions. This suggests that the spreadsheet modellers utilised simple tools to build complex models without utilising pre-defined structures.

When considering non programming error reduction methods, the significance of these findings for strategy formulation maybe substantial. Considering spreadsheets in terms of functionality may allow more accurate risk analysis which in turn could allow more effective application of controls to minimise such risks and improve quality. Indeed, further work should be conducted to crucially examine which structures within the functionality classes are the most prone to error. This data could then allow easier identification of spreadsheets that carry a higher risk since they are more error prone. This would then allow more informed decisions to be made in terms of risk management, spreadsheet use and auditing. One novel approach to reducing spreadsheet errors optimises on the use of logical operators (Thorne and Ball, 2005)

**A novel approach spreadsheet error management**

We have considered a novel approach to spreadsheet error management at UWIC (UK) called Example Driven Modelling (EDM). This requires the user to produce examples of attribute classifications (see below) which then deduce the function of those examples and generalises to new unseen examples. EDM uses machine learning techniques and research to date suggests this results in a more accurate spreadsheet. Machine learning, in the context of EDM, is best described as the ability to adapt and extrapolate patterns in data as defined by Russel and Norvig (2003). In particular, Neural Networks can be used in example attribute classifications of data. For example, the user provides simple examples of the problem data. This data is then fed into the learning machine and it produces an equivalent model of the problem. Thorne *et al.* (2004) discussed an experiment to test the relative levels of accuracy gained from both traditionally modelling a formulae and utilising an EDM approach, over successively more difficult problems. The results of this study found that producing the formulae with the traditional method was error prone (80% of models with error). The results of the EDM method yielded a much lower error rate (2% of models with error).

**Example Driven Modelling**

The basic premise of Example Driven Modelling is the concept that the user provides example data of the problem they wish to solve. The example data comprises of attribute classifications for the problem they wish to model. This example data set is then used to train a Neural Network and hence generalise the problem, see Figure 4.

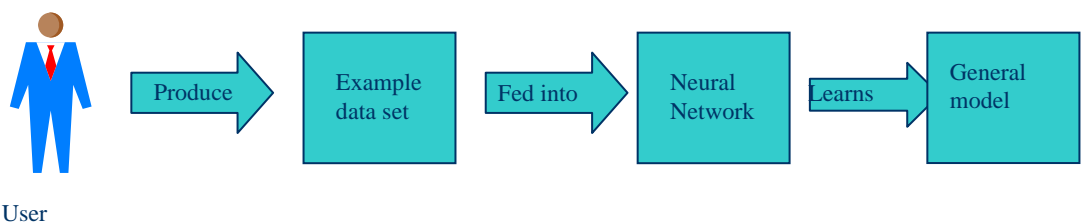

**Figure 4 Example Driven Modelling concept**

To demonstrate how this process works in practice, an example of EDM in execution will be presented. This example will extend from the construction of the data set to the performance results gained from the network after testing. The example problem is taken from an instruction book on how to implement Decision Support Systems (DSS) in Excel (Gross *et al.*, 2006). The example takes the form of a Credit Risk DSS that is entirely contained in one spreadsheet.

**Credit Risk Decision Support System**

This DSS is used to assess the credit worthiness of potential clients. The model classifies the applicants in to one of three possible classifications. These are: Accept; Further Enquire and Reject. These decisions are based upon equations that evaluate the classification based upon a number of variable inputs. These variable inputs are used as key identifiers as to the businesses worthiness for credit. These variables include: Current Year's sales, previous debt balances, Net worth and a number of risk class indexes.

Each classification has an associated rule that it must pass to satisfy the classification. For example, Rule 1 demands that Input 1 (Previous debt balances) is less than or equal to 10% of Input 2 (Current Year's Sales). If all of the conditions are True, the applicant passes on to the next rule for assessment. If the rule Fails, the applicant is given class 3 and classification equals reject. The rules increase in complexity for each class, they require more variables with more complex relationships to be satisfied. All of the rules and variable conditions are presented in Figure 5.

## Credit Risk Classification Flow Chart

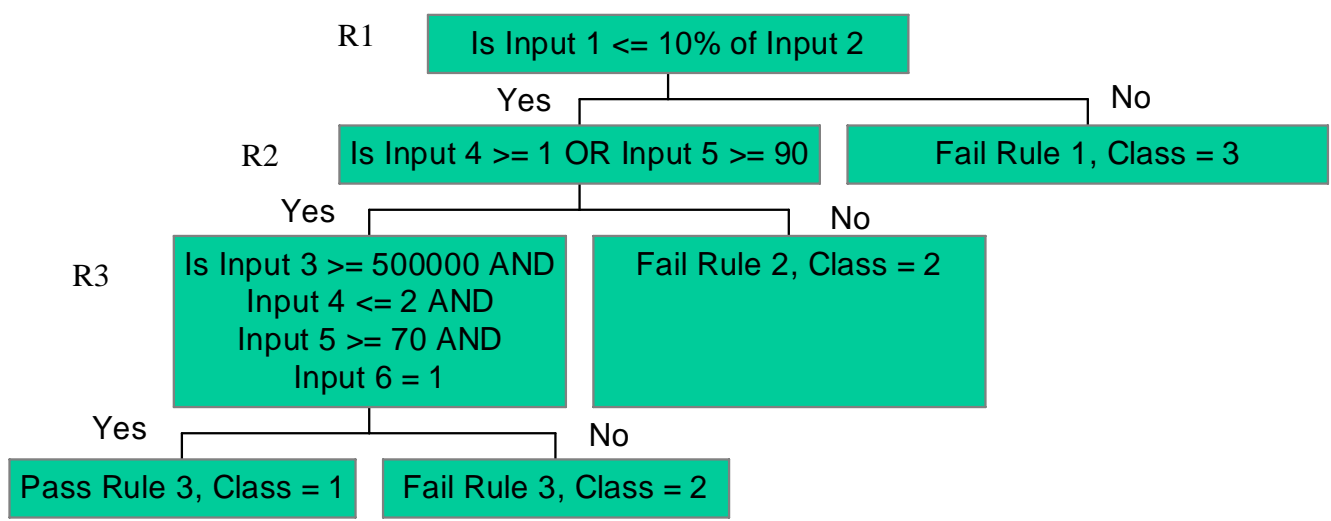

**Figure 5 Example risk classification system, generated from Gross *et al.* (2006)**

When using EDM the first task that must be completed is the generation of the example data set. This data set must cover all rules and all conditions that make up these rules. The simplest way to do this is to examine each rule in turn and consider what values satisfy or reject the rule's classification. Once the parameters of the model have been defined an example data set can be constructed around the true and false classifications of the model. The example data set appears as a set of values needed to satisfy or decline classifications in the model.

For example, if we examine Rule 3 (Classification 1) in Figure 4 it has the following conditions to be true:

*To satisfy Class 1, it is necessary that Variable 3 (Net worth) is greater than or equal to 50,000 AND Variable 4 (D&B Credit Index) is greater than or equal to 2 AND Variable 5 (D&B Paydex index) is greater than or equal to 70 AND Variable 6 (D&B Stress class index) is equal to 1.*

Based upon that statement the user must construct attribute classifications that both satisfy and reject that rule. For example, the user could construct an example where the Net Worth value equals 66,000, satisfying the first part of the rule. They would also need an example where Net Worth fails the rule i.e. a value of below 50,000 which would reject this part of the rule and the classification as a whole. This is of course only part of the rule, there are other clauses which need to have the same treatment. The complete set would comprise of a case of example values with the appropriate classifications that would be awarded by the model.

Once the example data set is constructed these values are then fed into a neural network so that the network can learn the dimensions of the problem based upon the parameters provided in the data set. An excerpt from the example data set used in this problem can be seen in Table 3, see appendix.

|  | CYS | PDB | NW | D&B C | D&B P | D&B S | C1 |
|---|---|---|---|---|---|---|---|
| Coded Var / example No. | Input 1 | Input 2 | Input 3 | Input 4 | Input 5 | Input 6 | Classification |
| 1 | 11000 | 500 | 50000 | 4 | 15 | 3 | 2 |
| 2 | 10000 | 2000 | 45000 | 3 | 20 | 3 | 1 |
|  |  |  |  |  |  |  |  |
| 3 | 27000 | 1000 | 75000 | 2 | 70 | 2 | 2 |
| 4 | 30000 | 2000 | 85000 | 1 | 85 | 2 | 2 |
| 5 | 45000 | 5000 | 69000 | 2 | 92 | 2 | 2 |
| 6 | 31000 | 500 | 77000 | 1 | 96 | 2 | 2 |
|  |  |  |  |  |  |  |  |
| 7 | 180000 | 2000 | 500000 | 2 | 79 | 1 | 3 |
| 8 | 210000 | 460 | 450000 | 1 | 72 | 1 | 2 |
| 9 | 100000 | 5000 | 505000 | 3 | 76 | 1 | 2 |
| 10 | 600000 | 2000 | 700000 | 1 | 69 | 1 | 2 |

**Table 3 Example Training Set**

**Training, testing and blind sets in Neural Networks**

During training and testing there are three fundamental components that allow the user to train and test a network accurately. The training set is used by the network as the source to learn the given problem. A training set, consisting of examples of input data for which the output is known, is presented to the network. The test set consists of examples not used in training but available to the network during training for cross validation. The blind set consists of examples that are completely unseen to the network and used in part to determine overall accuracy of the network given the universe of possibilities. The universe contains all possible training, test and blind sets.

**Results**

Once the model has learnt the problem, it outputs the Mean Squared Error (MSE) of the network (Chi squared). This value indicates how well the model has learnt the task and hence how well it will perform in testing. MSE indicates the difference between the training set and the actual output, i.e. a comparison between the network outputs and the known outputs.

Blind testing is the best absolute test, i.e. passing unseen data through the network and checking the classifications it gives on that basis. The trained network is given new examples and is assessed on how well it classifies those examples. Below in Figure 6 the results of the blind testing are displayed. In this blind test 25 unseen examples were passed through the trained network. The network values were then compared with the actual classifications that would be output given the input pattern. As can be observed, the network in this instance gave a high level of accuracy with no misclassifications. There are some values that diverge from the actual classifications but these values fall within the acceptable class value range that stop it from being misclassified.

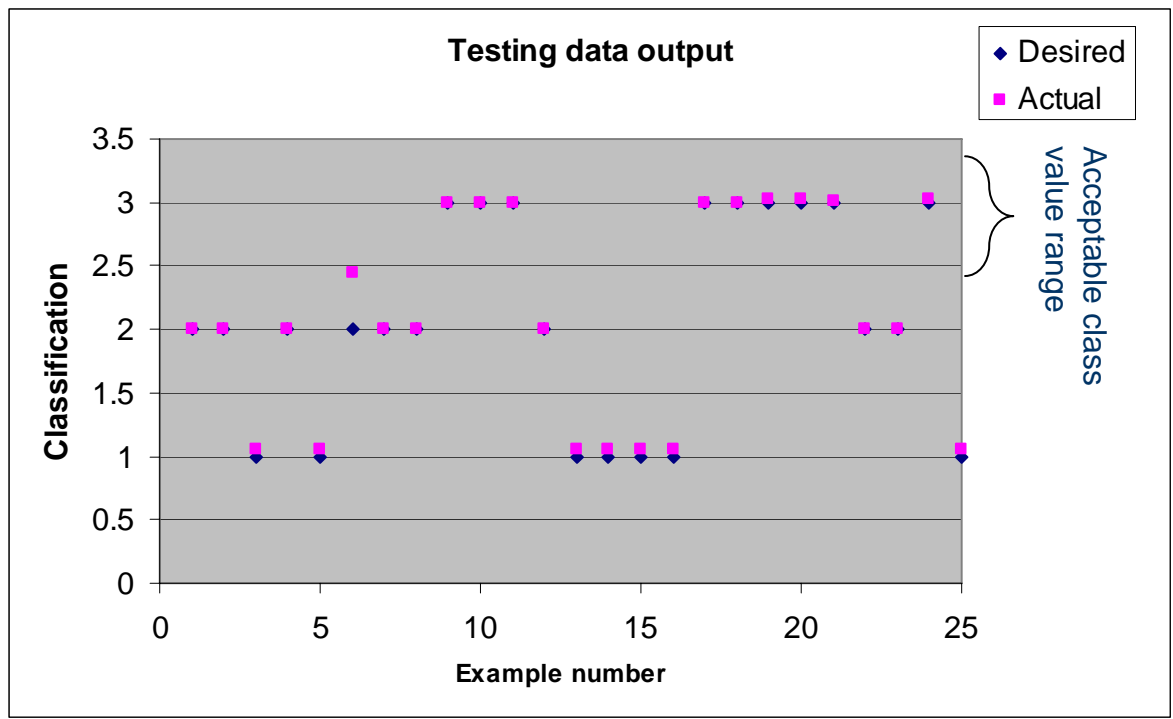

**Figure 6 Blind testing results**

**Conclusions**

The limitations of current spreadsheet research suggest that a deeper understanding of the use of operators and the operators relationship to error would provide fresh insight into the spreadsheet error problem.

The proposed method (EDM) performs well in business problems that are based upon logic in decision making. Beyond this domain, EDMs usefulness diminishes; it really requires some sort of decision process that includes a final conclusion.
In the presented example, the accuracy of the network was more than satisfactory, judging from the blind testing results, this model is unlikely to misclassify any examples presented to it.

Research shows (Thorne *et al.,* 2004) that although the process of generating examples is a very novel approach for users, it is actually easier than creating the equivalent formulae by programming a spreadsheet.

A weakness of EDM is that the model must be understood completely for the technique to be implemented effectively. However, in compensation, generating typical examples in normally a very easy task. Further research is required but ( as presented above) EDM appears to be more reliable than conventional spreadsheet methods particularly for the spreadsheet application subset:  decision support systems. Where domain knowledge is less sure the value of EDM maybe uncertain. Domain knowledge is often the Achilles heel of automated spreadsheet tools.

There may well be other challenges facing users of such a novel approach. In particular Base Error Rate (BER) could be a serious problem for users wishing to adopt this technique. The research shows that BER is most prevalent when the user is completing simple tasks in large volume. An example often quoted is copy typing, where the subject has to copy large amounts of text. This process yields a certain BER since it is repetitive and contains significant volume. Since the process of modelling using

EDM requires the user to create repetitive data sets based upon the rules of a problem, EDM may be at risk from a significant BER level.

This paper has demonstrated that a novel approach that has an appreciation for Human Factors can be executed successfully to replace a DSS implemented in spreadsheets.

**APPENDIX            Learning in Neural Networks**

There are a number of introductory texts on Neural Networks (NN), (Haykin 1999, Rumelhart *et al.* 1988, Principe *et al.* 1999). Briefly the procedure is as follows. A processing unit takes a number of input signals, $x_1,...,x_n$ with corresponding weights $w_1...,w_n$, respectively. These values are passed through the network to give an output which is then compared to the training set provided by the user. The network then adjusts the weights in an attempt to mimic the input/output pattern of the training set. This process is repeated until the network reaches some predetermined level of accuracy. This allows the network to become more and more accurate and hence the network learns the problem. The neuron will only be fired if the threshold function (T) is satisfied and is governed by this equation:

$$X_1W_1 + X_2W_2 + \ldots + X_nW_n > T$$

There are many different paradigms and algorithms for learning in neural networks. The most common is Backpropogation (supervised learning) since this offers the greatest generality (Haykin, 1999). The process of supervised learning follows the subsequent sequence:
1. A training set, consisting of examples of input data for which the output is known, is presented to the network.
2. The network weights are adjusted until the network produces results that are in agreement with the training set.
3. The network can then generalise to unseen examples in the universe

In addition to the backpropogation rule, genetic Optimisation is used in the input space for this example. GO is used in conjunction with NN to optimise the "Input space" of the problem. This has proved particularly useful when there is a limited amount of data available for training (Chang and Lippmann, 1990). The Genetic Algorithm seeks the best combination of inputs, i.e. the combination that gives the best accuracy.

**References**

Ayalew, Y., M. Clermont, R. Mittermeir. (2000), *'Detecting Errors in Spreadsheets'*, Proceedings of EuSpRIG Symposium, EuSpRIG 2000 Symposium, Spreadsheet Risks—the Hidden Corporate Gamble. (pp. 65-76). Greenwich, England: Greenwich University Press.

Ballinger. D, Biddle. R, Noble. P, (2003), *'Spreadsheet structure inspection using low level access and visualisation'*, Proceedings of the Fourth Australian user interface conference on User interfaces 2003, p.91-94, February 01, 2003, Adelaide, Australia

Brancheau. J and Wetherbe. J, (1990), *'The adoption of spreadsheet software: Testing Innovation Diffusion Theory in the context of End User Development'*, Information systems research, 1 (2), pp 115-142

Brown. P and Gould. J, (1989), *'An experimental study of people creating spreadsheets'*, ACM transactions on information systems, 5 (3), pp253-272

Chan and Storey (1996), *'The use of spreadsheets in organisations: determinates and consequences'*, Information and Management, 31, pp 119-134

Chang, E.I. and Lippmann, R. P. (1990), *'Using Genetic Algorithms to improve pattern classification performance'*, NIPS 1990, pp 797-803.

Croll. G, (2005), *'The importance and criticality of spreadsheets in the city of London'*, Proceedings of EUSPRIG 2005 – Managing spreadsheets in the light of Sarbanes-Oxley, London, UK, pp 82-94, ISBN 1-902724-16-X

Davis. B, (1987), 'Commentary on Information systems', Accounting horizons, 43

Fernandez, K. (2003), 'Investigation and Management of End User Computing Risk', Not published, MSc thesis April 2003. Available from University of Wales Institute Cardiff (UWIC) Business School

Gosling. C, (2003), 'To what extent are systems design and development techniques used in the production of non clinical corporate spreadsheets at a large NHS trust', Not published, MBA thesis May 2003, available from: University of Wales Institute Cardiff (UWIC) Business School.

Gross. D, Akaiwa. F, Nordquist. K, (2006), *'Succeeding in business with Microsoft Office: Excel 2003 a problem solving approach'*, Thompson Course Technology, Canada, ISBN 0-619-267-40-2, Page 254

Haykin. S, (1999), *'Neural Networks a comprehensive foundation'*, Prentice Hall publishers, 2nd Edition, New York, ISBN 0-13-908385-5

Hicks and Panko, (1995), 'Capital Budgeting Spreadsheet Code Inspection at NYNEX', Internet http://panko.cba.hawaii.edu/ssr/Hicks/HICKS.HTM, 12.1.05, 12.00, Available.

Janvrin. D and Morrison. J, (1996), 'Factors Influencing Risks and Outcomes in End-User Development' Proceedings of the Twenty-Ninth Hawaii International Conference on Systems Sciences, Vol. II, Hawaii, IEEE Computer Society Press, pp. 346-355.

Janvrin. D and Morrison. J, (2000), 'Using a structured design approach to reduce risks in End User Spreadsheet development', Information & management, 37, pp 1-12

Jenne. S, (1996), *Audits of End User Computing'*, Internal Auditor, 53 (6), pp 30-35

KPMG, (1997), 'Supporting the Decision Maker - A Guide to the Value of Business Modeling' press release, July 30, 1997.
http://www.kpmg.co.uk/uk/services/manage/press/970605a.html, Unavailable

Lawson. B, (2006), Private Communication with Barry Lawson of the Tuck Management School of Dartmouth College in the US, as Figure 3.

Napier. A. Batsell. R. Lane. D. Guadagno. N., (1992), 'Knowledge of command usage in a spreadsheet program', Database, 23 (1), pp 13-21.

Napier. A. Lane. D. Batsell. R. Guadango. N., (1989), 'Impact of restricted natural language interface on ease of learning and productivity', Communications of the ACM, 32 (10), pp 1190-1198

Panko. R and Halverson. R, (1997), ' Are Two Heads Better than One? (At Reducing Errors in Spreadsheet Modelling?' Office Systems Research Journal, 15 (1), pp. 21-32.

Pemberton. J and Robson. A, (2000), *'Spreadsheets in business'*, Industrial management and systems, 100 (8), pp 379-388
Principe. J.C., Euliano. N. R., Lefebvre. W.C., (1999). *'Neural and adaptive systems: Fundamentals through simulation'*, John Wiley and Sons, New York, ISBN 0-471-35167-9

Rajalingham K.; Chadwick, D.; Knight, B. & Edwards, D., (2000), *'Quality Control in Spreadsheets: A Software Engineering-Based Approach to Spreadsheet Development'*, Proceedings of the 33rd Hawaii International Conference on System Sciences, Maui, Hawaii.

Rajalingham. K., (2005), *'A revised classification of spreadsheet errors'*, Proceedings of EUSPRIG 2005 – Managing spreadsheets in the light of Sarbanes-Oxley, London, UK, pp 185-200, ISBN ?

Rumelhart. D.E., McCelland. J.L., (1988), *'Parallel distributed Processing'*, The MIT Press, Cambridge Massachusetts, ISBN 0262-18120-7

Russel. S and Norvig. P, (2003), *'Artificial Intelligence – A Modern Approach'*, 2nd Edition, Pearson education inc., New Jersey, ISBN 0-13-080302-2

SERP, (2005), *'Spreadsheet Engineering Research Project'*, Tuck Business School, Dartmouth College, USA

Taylor. M, Moynihan. P, Wood-Harper. T, (1998), 'End User Computing and information systems methodologies', Information systems Journal, 8, pp 85-96

Teo.T and Tan. M, (1999), *'Spreadsheet development and 'what if' analysis, quantitative versus qualitative error'*, Accounting management and Information Technology, 9 (3), pp 141-160
Thorne. S and Ball. D, (2005), *'The relevance of Human Factors to the management of spreadsheet risks'*, INFORMS 2005, San Francisco

Thorne. S. Ball. D. Lawson. Z., (2004), 'A novel approach to spreadsheet formulae production and overconfidence measurement to reduce risk in spreadsheet modelling', Proceedings of EUSPRIG 2004 – Risk reduction in End User Computing, Klagenfurt, pp 71-85, ISBN 1 902724 94 1